\documentstyle[12pt]{article}
 1
\title{\marginpar{\vspace{-1in}\hspace{-1in}\small\begin{tabular}{l}
KFT U{\L} 7/95\\ hep-ph/9506277\end{tabular}}
DOMINATING CONSTITUENT \protect\\
 OF THE $\Theta$ MESON}
\author{M. Majewski\thanks{{\em E-mail:}\,{\tt mimajew@mvii.uni.lodz.pl}
}\vspace{2ex}\\
University of {\L}\'od\'z\\
Departament of Theoretical Physics\\
ul.~Pomorska 149/153\\
90-236 {\L}\'od\'z, Poland }
\date{June 1995}
\begin{document}
\maketitle
\thispagestyle{empty}
\vfill
\begin{abstract}
On the basis of  the  relation  between  values  of  the
coupling constants determined from data fit it is concluded
that the dominating constituent  of  the  $\Theta$  meson  is  the
glueball. The possibility that  $\Theta$  meson  decays  into  two
$\sigma(750)$ mesons is suggested.
\end{abstract}
\vfill
\newpage
\setcounter{page}{2}
\section{Introduction}
\indent

$\Theta$(1720) meson,  since  its  first  observation  [1]  was
regarded as a glueball candidate with spin-parity $J^{PC}= 2^{++}$  .
Later its glueball assignment  was  put  in  question.  The
reason was that  Lattice  QCD  (see  e.g.[2])  and  several
phenomenological models [3] consistently predicted the mass
of the lightest tensor glueball to be about 2 times  bigger
than the mass of the lightest  scalar  glueball,  while  an
appropriate scalar candidate was lacking; the mass  of  the
only serious scalar glueball candidate G(1590) [4]  is  far
too high for that. As  it  is  difficult  to  find  another
interpretation for the nature of the $\Theta$ meson, its spin  was
questioned: in one measurement [5] it has  been  determined
as $0$. Such assignment simply shifts the difficulty from the
tensor multiplet to the much more  mysterious  scalar  one
and rather stops then stimulates the investigation of the $\Theta$
meson and the glueball search. Later it was argued [6] that
the spin of the $\Theta$ meson is not yet firmly  established  and
now it is considered to be $0$ or $2$ [7].

There are several reasons to reinvestigate  the  problem
of the $\Theta$ meson, assuming $J^{PC}=2^{++}$   for it.

First in  a  recent  experiment   [8]   a   new   scalar
glueball candidate  has  been  discovered.  Let's  call  it
$\sigma$(750). If $\sigma$ and $\Theta$  are  both  glueballs,
  then  the  right
hierarchy of the glueball masses is restituted and there is
no reason to contest the spin $2$ for $\Theta$. Moreover, on account
of the rate of their masses the reaction $\Theta\Rightarrow\sigma\sigma$
 is possible
and since it is allowed, the $\sigma$ mesons should  be  copiously
produced in the $\Theta$ decays. This  enables  us,  possibly,  to
solve not  only  the  controversial  problem  of  the  very
existence of the $\sigma$ meson [7], but also the problem  of  the
nature of both these mesons. Although this may be only done
experimentally, room for such decay channel may be found by
comparison of the $\Gamma^{\rm tot}(\Theta)$ with the sum
of its  partial  two
$q\bar q$ meson widths.

Second we notice that hitherto analyses of the $\Theta$ problem
[9,10,11,12,13]  are  inconsistent.  They   require   exact
realization of the OZI prohibition  for  the  quarkonium  -
quarkonium transitions but not require it for the  glueball
- quarkonium ones which are also  forbidden.  Consequently,
they use two coupling constants to  describe  data  on  the
$f_2,f'_2,\Theta$  decays  into  pseudoscalar  mesons,   while   the
consistent description requires the three ones. Fitting the
three coupling constants to the data we should be  able  to
determine coupling constants for different  OZI  suppressed
reactions and verify agreement between  them.

\eject
\noindent
It  will  be
also seen that the known difficulty with smallness  of  the
predicted $f_2\Rightarrow\pi\pi$ width [10,11]
does not appear in  the  consistent
description.

Third there is a problem of the  mixing  model.  Looking
for any nonexotic  glueball  we  have  to  use  some  model
describing mixing of the  isoscalar $q\bar q$  states  with  pure
glueball state $G$. The result of the analysis depends on the
mixing model. During $70^{\rm th}$   and early $80^{\rm th}$
a  number  of  the
models has been formulated [9,10,12,14,15,16,17].  However,
it is still unclear whether any of  them  is  adequate  and
whether the same model will work in  different  cases.  The
most favoured was gluon exchange model (GEM) based  on  QCD
inspired mass operator [18,16]. This model has been applied
to describe mixing of  the  $\imath$(1440)  meson  with  the  $\eta,\eta'$
mesons and (what is rather natural) completely failed [19].
Also it has been claimed unsatisfactory in  an  attempt  to
solve the problem of the $\Theta$ meson. So  it  remained  unclear
whether the  gluon  exchange  model  is  adequate  or  not.
It
is thus reasonable to look for further  mixing  models  and
verify their predictions. We test the model which was  for\-mulated
later than the models mentioned  above  [20,21,22].
The model has great predictive power  and  is  transparent.
Its transparency is mainly due to the set  of  inequalities
restricting masses of the mixing particles. For example, it
follows from these restrictions that mesons $\omega$ and $\phi$  cannot
include the admixture of the state having higher  mass,  it
also follows that the meson $\imath$(1440) cannot have  admixtures
of the $\eta$  and  $\eta'$  mesons.  These  qualitative  predictions
agree very well with data, as the  vector  glueball  is not
observed (and even not expected) in the region above 1 GeV;
also the $\imath$(1440) meson should mix first  of  all  with  the
states from its nearest neighborhood which  do  exist  [7].
The quantitative predictions of the model can only be  made
and verified for $2^{++}$   mesons, as they  constitute  the  only
well known multiplet obeying mixing conditions. It will  be
seen below that predictions of the model agree  well  with  data.
Here we merely notice that this investigation  considerably
vindicates GEM. Namely, GEM would give the same results, if
the masses of the physical mesons were changed  within  1---2
standard  deviations.  However  GEM  does  not   give   any
suggestion how the masses should  be  changed.  We  do  not
discuss this problem in detail in the present paper.

\eject

\section{Notations and results}
\indent

2.1 In a number of papers [23,20,21,24,22] a model  for
mixing the $q\bar q$ with any SU(3) singlet has  been  formulated.
The isoscalar physical states of  the  decuplet  of  mesons
formed in such a way may be expressed by  the  ideal  quark
states
\begin{equation}
N=\frac{1}{\sqrt{2}}(u\bar u+d\bar d),\hspace{2cm} S=s\bar s
\end{equation}

\noindent and an additional state $G$ which may be, in
principle, any SU(3) singlet. We have
\begin{equation}
\left[\begin{array}{c}
f_2\\
f'_2\\
\Theta
\end{array}\right]=V\left[\begin{array}{c}
N\\
S\\
G
\end{array}\right]
\end{equation}

\noindent where  the  mixing  matrix  V  does  not  depend  on   free
parameters and is completely determined by  masses  of  the
mesons belonging to the decuplet. As there are $5$  different
masses, the dependence on them is, in general, complicated,
but for the decuplet of mesons  $a_2 ,K_2 ,f_2 ,f'_2,\Theta$  it  becomes
simple: the elements of the matrix V depend mainly  on  one
variable $\Delta$ which may be chosen as the difference between $a_2$
and $f_2$  meson masses [22]
\begin{equation}
\Delta=m_{a_2}-m_{f_2},
\end{equation}

\noindent while their dependence on  other  mass  variables  is  weak
under reasonable mass variations.

   We use the following notation for the  elements  of  the
matrix V
\begin{equation}
V=\left[\begin{array}{ccc}
x_1&y_1&u_1\\
x_2&y_2&u_2\\
x_3&y_3&u_3
\end{array}\right].
\end{equation}
\vspace{1cm}

2.2 The relative radiative widths are  calculated  from
the formula [10]

\begin{equation}
\frac{\Gamma_{\gamma\gamma}(f^{(j)}_2)}{\Gamma_{\gamma\gamma}(a_2)}=
{\rm ph.space}\times(x_j+\frac{\sqrt{2}}{5}y_j)^2
\end{equation}

\noindent where j=1,2,3 correspond to $f_2 ,f_2',\Theta$
mesons,  respectively.
Predictions for the radiative widths depend exclusively  on
$\Delta$ (3). Therefore the data  on  these  widths  control  this
quantity. According to present data [7] only for
$f_2'\Rightarrow\gamma\gamma$
width there is a disagreement. The predicted width  is  too
small $(\Gamma_{\gamma\gamma}(f_2')\simeq 0.1 {\rm keV})$ for
$\Delta= 44 {\rm MeV}$ corresponding  to
the mean experimental masses of the $a_2$
and $f_2$   mesons  (c.f.  [10]).  We
increase $\Gamma_{\gamma\gamma}(f_2')$ choosing the masses such
that $\Delta = 34 {\rm MeV}$.
As  an  input  for  further  calculations  we  choose   the
following values of the masses (already satisfying the mass
formula):
\begin{equation}
\begin{array}{ccc}
m_{a_2}=1.316{\rm GeV},&m_{K_2}=1.432{\rm GeV}& \\
m_{f_2}=1.282{\rm GeV},&m_{f_2'}=1.516{\rm GeV},&m_\Theta=1.720{\rm GeV}.
\end{array}
\end{equation}

\noindent We get following mixing matrix

\begin{equation}
V=\left[\begin{array}{ccc}
0.9585&0.0825&0.2730\\
-0.1679&0.9370&0.3064\\
-0.2306&-0.3395&0.9119
\end{array}\right],
\end{equation}

\noindent and the rates of the radiative widths

\begin{equation}
\frac{\Gamma_{\gamma\gamma}(f_2^{(j)})}{\Gamma_{\gamma\gamma}(a_2)}=
\left\{\begin{array}{c}
2.48\\
0.04\\
0.66
\end{array}
\right. .
\end{equation}

\noindent Notice that for the nonet mixing ($\Theta$ is absent) these  rates
are

\begin{equation}
\frac{\Gamma_{\gamma\gamma}(f_2^{(j)})}{\Gamma_{\gamma\gamma}(a_2)}=
\left\{\begin{array}{c}
2.64\\
0.13
\end{array}.
\right.
\end{equation}
\vspace{1cm}

2.3 The  widths  of  the  strong  two-body  decays   are
calculated using the formula [25]

\begin{equation}
\Gamma_{mn}(k)=\frac{p^5}{M_k}g^2_{kmn},
\end{equation}

\noindent where $k$ is the decaying particle; $m,n$ are  decay  products;
$M_k$  is the mass of the decaying particle; $p$ is c.m. momentum
of the decay product; $g_{kmn}$    are quantities depending on  the
coupling constants and contents of the  $N,S,G$  states.  For
$2^+\Rightarrow1^- 0^-$   decays  $g_{kmn}$     are  expressed
by  one  coupling
constant $g_V$ . For $2^+\Rightarrow 0^- 0^-$
decays  they  are  expressed  by
three coupling  constants  $g_8 ,g_0 ,g_G$   corresponding  to  the
$(q\bar q)_{\rm octet}$ , $(q\bar q)_{\rm singlet}$ and $G$ states.
Choosing  de'Swart
phases we get for the  particle  widths  of  the  isoscalar
mesons the following expressions

\begin{eqnarray*}
\lefteqn{\Gamma_{\pi\pi}(f_2^{(j)})=}\\
&&=\frac{3}{2}\frac{p^5}{M_j}\left[
\frac{1}{\sqrt{12}}(g_0-\frac{2}{\sqrt{5}}g_8)x_j+\frac{1}{\sqrt{24}}
(g_0+\frac{4}{\sqrt{5}}g_8)y_j+\frac{1}{\sqrt{8}}g_Gu_j\right]^2,
\end{eqnarray*}

\begin{eqnarray}
\lefteqn{\Gamma_{K\bar K}(f_2^{(j)})=}\nonumber\\
&&=2\frac{p^5}{M_j}\left[
\frac{1}{\sqrt{12}}(g_0+\frac{1}{\sqrt{5}}g_8)x_j+\frac{1}{\sqrt{24}}
(g_0-\frac{2}{\sqrt{5}}g_8)y_j+\frac{1}{\sqrt{8}}g_Gu_j\right]^2,
\end{eqnarray}

\begin{eqnarray*}
\lefteqn{\Gamma_{\eta\eta}(f_2^{(j)})=}\\
&&=\frac{1}{2}\frac{p^5}{M_j}\left[
\frac{1}{\sqrt{12}}(g_0+\frac{2}{\sqrt{5}}g_8)x_j+\frac{1}{\sqrt{24}}
(g_0-\frac{4}{\sqrt{5}}g_8)y_j+\frac{1}{\sqrt{8}}g_Gu_j\right]^2{\cos^4\Theta}
\end{eqnarray*}

\noindent where $\Theta_P$  is mixing angle of  the
pseudoscalar  mesons  and
$x_j ,y_j ,z_j$  are elements of the mixing  matrix  (4).  Coupling
constants are determined in the following way:
$1^0$  $g_V$  is determined from the fit to
$a_2\Rightarrow\rho\pi$ partial width.
Other $2^+\Rightarrow 1^- 0^-$  widths, including the non yet measured ones
for $f_2',\Theta\Rightarrow\bar K^* K$ + c.c. are calculated.
$2^0$  $g_8 ,g_0 ,g_G$  are  determined  from  the  fit  to  the  width
$\Gamma_{\pi\pi}(f_2)$ and to the rates
\begin{equation}
\frac{\Gamma_{\pi\pi}(f_2^{(j)})}{\Gamma_{K\bar
K}(f_2^{(j)})+\Gamma_{K^*\bar K+K\bar K^*}(f_2^{(j)})}
\end{equation}
for $f_2^{(j)}$ equal  $f_2'$ and $\Theta$.
We  add  to  the  $f_2',\Theta\Rightarrow K\bar K$
widths the $f_2',\Theta\Rightarrow\bar K^* K$ + c.c.
ones, as the latter  reactions
are not separately recorded.

The results of the calculation and the fitted values  of
the coupling constants are  given  in  the  Table.  Similar
predictions for nonet mixing and corresponding experimental
values are also given. To determine the $g_8$  and  $g_0$   in  the
nonet case we fit $\Gamma_{K\pi}(K_2)$ and the rate (12) for $f_2'$.

\eject

\section{Discussion}
\indent

3.1 The model gives us predictions of the three
kinds:
\begin{description}
\item[-] mass formula enabling us to calculate one of the masses
\item[-] rates of the radiative width testing  difference  between
two of the masses
\item[-] two-body strong decay widths.
\end{description}

It follows from the mass formula that, in order to obey it,
we must slightly shift the masses from  their  mean  values
(both for decuplet and the nonet). We also find  that  only
one of the radiative widths, the
$\Gamma_{\gamma\gamma}(f_2')$ one,  poses  a
problem. It is too high in the nonet pattern and too low in
the decuplet one. In the latter case  we  must  reduce  the
mass difference $\Delta$ to increase
$\Gamma_{\gamma\gamma}(f_2')$.  As  a  result  of
these two modifications we shift the masses in  the  decuplet  pattern
within two standard deviations. Although for the
nonet the mass formula is satisfied better,  the  agreement
for the nonet $\Gamma_{\pi\pi}(f_2)$ is worse, as the calculated value  is
rather too high. Other $2^+\Rightarrow 0^- 0^-$  widths are described  well
both in the decuplet and the nonet, with exception of
$a_2\Rightarrow\eta\pi$
and $K_2\Rightarrow\eta K$ ones (c.f. [9]). The
$2^+\Rightarrow1^- 0^-$ widths  are
described  well in both multiplets, except the
$K_2\rightarrow K^*\pi$ one which is too high (c.f. [11]).
\vspace{1cm}

3.2 Having  determined  $g_8 ,g_0 ,g_G$   we  can  calculate  the
coupling constants of the OZI suppressed decay
$S\Rightarrow\pi\pi$

\begin{equation}
g_{S\pi\pi}=\frac{1}{\sqrt{24}}(g_0+\frac{4}{\sqrt{5}}g_8)=0.112.
\end{equation}

\noindent Comparing it with the coupling constant of the
$G\Rightarrow\pi\pi$ decay

\begin{equation}
g_{G\pi\pi}=\frac{1}{\sqrt{8}}g_G=0.160,
\end{equation}

\noindent we find that $g_{S\pi\pi}$    and $g_{G\pi\pi}$
have the same order of magnitude

\begin{equation}
g_{G\pi\pi}\cong1.43 g_{S\pi\pi}.
\end{equation}

\noindent This indicates that transition between G state
and two pion
(or any two  pseudoscalar  meson) state is  suppressed  by  OZI
mechanism. Therefore eq. (15) gives the evidence that $G$  is
the glueball state.

If we also define the coupling constant of the  $N\Rightarrow\pi\pi$
transition $g_{N\pi\pi}$    as
\begin{equation}
g_{N\pi\pi}=\frac{1}{\sqrt{12}}(g_0-\frac{2}{\sqrt{5}}g_8)=1.172,
\end{equation}

\noindent we can introduce the OZI suppression factor $\gamma_{\rm OZI}$

\begin{equation}
\gamma_{\rm
OZI}=\frac{g_{S\pi\pi}}{\frac{1}{\sqrt{2}}g_{N\pi\pi}}=0.135.
\end{equation}

Observe that coupling constants $g_8 ,g_0 ,g_G$,  as well as  the
mixing matrix $V$, are functions of only one mass variable  $\Delta$.
It  has  been  examined  by  multiple  fit  that  $g_{N\pi\pi}$      is
insensitive to $\Delta$, $g_{S\pi\pi}$ and $g_{G\pi\pi}$
depend on  $\Delta$  approximately
linearly and eq. (15) is satisfied for 12 MeV $\leq \Delta\leq$ 40  MeV
with good accuracy.

For the nonet we find
\begin{displaymath}
[g_{N\pi\pi}]_{\rm non}=1.261,
\end{displaymath}
\begin{equation}
[g_{S\pi\pi}]_{\rm non}=0.191.
\end{equation}

Smallness of the $G$ state content and of the  $g_{G\pi\pi}$     value
explains weak influence of the glueball state on the widths
of the $f_2$  and $f_2'$ mesons. To choose on this ground  between
the nonet and decuplet pattern (if it is possible  at  all)
more accurate data are required (especially the data on $f_{2}'
\Rightarrow K^* \bar{K}+\bar{K}^* K$ are lacking).
 However, it should be noticed that the value of  the
constant $g_G$  is determined mainly by the  data  on  $\Theta$  meson
(eq. (12) for $\Theta$) and consequently the weak influence of the
state $G$ on the  $f_2$ ,  $f_2'$  states  is  consistent  with  the
information on $\Theta$. On the  other  hand,  to  understand  the
properties of the $\Theta$ meson, the mixing of the  state  $G$  with
the states $N,S$ is necessary. For example,  small  value  of
the $\Gamma_{\pi\pi}(\Theta)$ follows not only from
smallness of the $g_{G\pi\pi}$,
but
also from destructive interference of the states $G$ and $N$.
\vspace{1cm}

3.3 Due to the smallness of the $g_{G\pi\pi}$   ,
the sum of the  partial
$\Theta$ widths over two-body channels is also small

\begin{equation}
\Gamma_{\pi\pi}+\Gamma_{K\bar
K}+\Gamma_{\eta\eta}+\Gamma_{\bar KK^*+\bar K^* K}=32.8
\end{equation}

\noindent This should be compared with the total width $\Gamma^{\rm
tot}(\Theta)=140{\rm MeV}$
(see, however, [26]). So large disagreement needs  some
interpretation. Two possible  explanations,  not  excluding
each other, seem to be the most plausible.
\begin{description}
\item[$1^0$]  The observed $\Gamma^{\rm tot}(\Theta)$
includes part of  the  signal  from
scalar meson which do exists in this mass region [4].
\item[$2^0$]  The sum (19) does not include the main decay channel
\end{description}
\begin{equation}
\Theta=2\sigma(750).
\end{equation}

   The second explanation  looks  especially  exciting.  It
suggests measurement of the decay channel
$\Theta\Rightarrow4\pi$ in looking
for the reaction (20) which should dominate  the  $\Theta$  decay.
Assuming 100 MeV for its partial width, we find $g_{G\sigma\sigma}\cong3.6$.
Confirmation of this reaction would be important  not  only
for understanding the nature of the $\Theta$ and $\sigma$ mesons, but for
glueball search and the strong interaction physics as well.
\vspace{1cm}

   Author thanks profs.  V.A.Meshcheryakov,  S.B.Gerasimov,
W.Tybor and  P.Kosinski  for  interest  to  this  work  and
valuable comments.


\newpage

\hoffset -3cm
\hsize18cm

Table. Hadron decays of tensor mesons\\[-0.5cm]
\begin{flushleft}
\begin{tabular}{cccccccc}
\hline
Particle&&\multicolumn{2}{c}{Decay Mode}&
\multicolumn{4}{c}{Width}\\
\cline{5-8}
&$\begin{array}{c}M\\[-0.1cm]\Gamma_{\rm tot}\end{array}$
& & &\multicolumn{2}{c}{Experim.}&
\multicolumn{2}{c}{Calculated (MeV)}\\
&(MeV)&$VP$&PP&$VP$&PP&Decuplet&Nonet\\
\hline
$a_2$& &$\Rightarrow\rho\pi$& & $(70.1\pm2.7)\%$& &$67.6^{\rm
inp}$&$68.6^{\rm inp}$\\
 &$1318.4\pm0.7$& &$\Rightarrow\bar K K$& &$(4.9\pm0.8)$\%&6.0&5.7\\
 &$102.7\pm2.2$& &$\Rightarrow\eta\pi$& &$(14.5\pm1.2)$\%&11.0&10.5\\
  & & &$\Rightarrow\eta'\pi$& &$<1$\%&0.01&0.01\\[0.6cm]
$K_2$& &$\Rightarrow\rho K$& &$(8.7\pm0.8)$MeV& &9.5&9.5\\
 &$1425.4\pm1.3$&$\Rightarrow K^*\pi$& &$(24.8\pm1.7)$MeV& &32.1&32.1\\
 &$98.4\pm2.4$&$\Rightarrow\omega K$& &$(2.9\pm0.8)$MeV& &3.1&3.1\\
 & & &$\Rightarrow K\pi$& &$(48.9\pm1.7)$MeV&50.4&$47.6^{\rm inp}$\\
 & & &$\Rightarrow K\eta$& &$(0.14\pm
        \begin{array}{c}0.28\\[-0.1cm]0.09\end{array})$MeV&1.6&1.6\\[0.6cm]
$f_2$& & &$\Rightarrow\pi\pi$& &$(156.7\pm
        \begin{array}{c}3.0\\[-0.1cm]1.3\end{array})$MeV&$155.1^{\rm
inp}$&177.8\\
 &$1274\pm5$& &$\Rightarrow\bar KK$& &$(8.6\pm0.9)$MeV&9.7&11.9\\
 &$185\pm20$& &$\Rightarrow\eta\eta$& &$(0.83\pm0.19)$MeV&0.5&0.7\\[0.6cm]
$f_2'$& & &$\Rightarrow\pi\pi$& &$(0.70\pm0.14)$MeV&0.48&0.74\\
 &$1525\pm5$& &$\Rightarrow\bar KK$& &$(61\pm5)$MeV&41.7&55.0\\
 &$76\pm10$& &$\Rightarrow\eta\eta$& &$(23.9\pm
        \begin{array}{c}2.2\\[-0.1cm]1.3\end{array})$MeV&11.3&14.1\\
 & &\multicolumn{2}{c}{$\Rightarrow K^*\bar K+\bar K^*K$}& & &7.2&8.6\\
 & &\multicolumn{2}{c}{$\Rightarrow\pi\pi$}& & & &  \\
\cline{2-4}
 &\multicolumn{3}{c}{\raisebox{-2mm}{$\Rightarrow K\bar K+(K^*\bar K+
 \bar K^* K)$}}&
&\raisebox{3mm}[-0.6cm]{$0.0115\pm0.0022$}&
\raisebox{3mm}[-0.6cm]{$0.0097^{\rm inp}$}&
\raisebox{3mm}[-0.6cm]{$0.0117^{\rm inp}$}
\end{tabular}
\end{flushleft}
\eject

\hoffset -1cm
\hsize18cm

Table. Hadron decays of tensor mesons-continuation\\[-0.5cm]
\begin{flushleft}
\begin{tabular}{cccccccc}
\hline
Particle&&\multicolumn{2}{c}{Decay Mode}&
\multicolumn{4}{c}{Width}\\
\cline{5-8}
&$\begin{array}{c}M\\[-0.1cm]\Gamma_{\rm tot}\end{array}$
& & &\multicolumn{2}{c}{Experim.}&
\multicolumn{2}{c}{Calculated (MeV)}\\
&(MeV)&$VP$&PP&$VP$&PP&Decuplet&Nonet\\
\hline
$\Theta$& & &$\Rightarrow\pi\pi$& &$(3.90\pm
       \begin{array}{c}0.20\\[-0.1cm]2.40\end{array})$\%&10.0& \\
 &$1713.2\pm
       \begin{array}{c}1.9\\[-0.1cm]4.5\end{array}$& &$\Rightarrow\bar K K$&
       &$(38\pm
       \begin{array}{l}9\\[-0.1cm]19\end{array})$\%&16.6& \\
 &$138\pm
       \begin{array}{l}12\\[-0.1cm]9\end{array}$&
  &$\Rightarrow\eta\eta$& &$(18\pm
       \begin{array}{c}2.0\\[-0.1cm]13.0\end{array})$\%&3.8& \\
 & &\multicolumn{2}{c}{$\Rightarrow K^*\bar K+\bar K^*K$}& & &2.4& \\
 & &$\Rightarrow\pi\pi$& & & & & \\
\cline{2-4}
 &\multicolumn{3}{c}{\raisebox{-2mm}{$
\Rightarrow\bar K K+(K^*\bar K+\bar K^* K)$}}&
&\raisebox{3mm}[-0.6mm]{$0.39\pm0.14$}&
\raisebox{3mm}[-0.6mm]{$0.53^{\rm inp}$}& \\
\hline
\end{tabular}
\end{flushleft}

The values of the constants\\
\parbox[t]{8cm}{
Decuplet\\
Mass input:\\
$a_2=(1.316{\rm GeV})^2, K_2=(1.432{\rm GeV})^2,$\\
$f_2=(1.282{\rm GeV})^2, \Theta=(1.720{\rm GeV})^2;$\\
Coupling constants determined from fit\\
$g_V=1.862$, $g_8=-1.315$,\\
$g_0=2.885$, $g_G=0.4525$}
$\qquad$
\parbox[t]{7cm}{
Nonet\\
Mass input:\\
$a_2=(1.318{\rm GeV})^2$, $K=(1.432{\rm GeV})^2$, $f_2=(1.275{\rm GeV})^2,$\\
Coupling constants determined from fit\\
$g_V=1.862, g_8=-1.278$, $g_0=3.224$}\\[0.6cm]
For the decay products we assume:\\
$0^-: m_\pi=139.6{\rm MeV}, m_K=495.6{\rm MeV}, m_\eta=548.8{\rm MeV},
\Theta_P=-10^0$\\
$1^-: m_\rho=768.3{\rm MeV}, m_{K^*}=894.0{\rm MeV}, m_\omega=781.95{\rm
MeV}, \Theta_V=35^0$
\end{document}